\documentclass[showpacs,preprintnumbers,amsmath,amssymb,published]{JHEP3}
\usepackage{graphicx}
\usepackage{dcolumn}
\usepackage{bm}

\title{The flavour projection of staggered fermions and the quarter-root trick}
\author{Steven Watterson\\ Department of Mathematics, Trinity College, Dublin 2, Ireland and \\ Division of Pathway Medicine, University of Edinburgh Medical School, Chancellor's Building, 49 Little France Crescent, Edinburgh, EH16 4SB, Scotland\\ E-mail: \email{watterss@maths.tcd.ie}}

\abstract{
It is shown that the flavour projection of staggered fermions can be written as a projection between the fields on four separate, but parallel, lattices, where the fields on each are modified forms of the standard staggered fermion field.  Because the staggered Dirac operator acts equally on each lattice, it respects this flavour projection.  We show that the system can be gauged in the usual fashion and that this does not interfere with flavour projection.  We also consider the path integral, showing that, prior to flavour projection, it evaluates to the same form on each lattice and that this form is equal to that used in the quarter-root trick.  The flavour projection leaves a path integral for a single flavour of field on each lattice.}

\keywords{Lattice Quantum Field Theory, Lattice Gauge Field Theories, Lattice QCD}
\received{8th May, 2007}
\accepted{8th June, 2007}
\JHEP{06(2007)048}
\begin{document}

\section{Introduction}
Constructing lattice field theories that incorporate a single non-degenerate fermion field, with chiral symmetry, has proven to be very challenging.  Wilson originally identified the problem by showing that na\"{i}vely discretizing the continuum Dirac equation led to a doubling in the number of fermion fields in each dimension of momentum space \cite{Wilson}.  As a solution, he proposed adding a term proportional to the second derivative to break the degeneracy between the physical field and the doublers. However, this term broke the chiral symmetry of the theory.  Subsequently, Kogut, Susskind and Banks proposed the staggered fermion formulation \cite{KSB1}\cite{KSB2}\cite{KSB3} in which the number of degenerate fields is reduced to $2^{n/2}$ in $n$ dimensions and which retains a limited form of chiral symmetry.  Other formulations have since succeeded in isolating a single chiral field by introducing an infinite number of regulating fields \cite{Kaplan}\cite{Furman}\cite{Neuberger}.  In the finite case, these formulations contain an exponentially suppressed term that breaks the chiral symmetry of the theory.

For reasons of computational efficiency, the staggered fermion formulation has remained popular in lattice calculations and it is common to see the quarter-root trick used to model a system with one flavour of fermion \cite{Davies}\cite{Aubin}.  It has been shown that, in the free field case, it is possible to construct a lattice Dirac operator whose determinant is the quarter-root of the determinant of the conventional staggered Dirac operator \cite{Adams}.  Although, the locality of this operator has been questioned \cite{Bunk}, numerical studies suggest that a suitably local operator does exist \cite{Peardon}.

In this paper, we draw on previous work \cite{me} to show that the flavour projection of staggered fermions can be written as a projection between four modified staggered fermion fields on four separate, but parallel, lattices.  The staggered Dirac operator respects the flavour projection because it acts upon the fields on each lattice equally.  We show that we can gauge the fields in each lattice in the usual way and that the flavour projection is equally valid in the gauged case.  We discuss the role of chiral symmetry and we see that, in this framework, the $\gamma_5\otimes\gamma_5$ chiral symmetry of the staggered fermion formulation can be used to isolate the chiral components of the fields.  Finally, we consider the path integrals on each lattice, before and after flavour projection, showing that projection leaves the path integral for a single flavour of field on each lattice.  In considering this, we see that evaluating the path integral, prior to flavour projection, leads to an expression identical to that used in the quarter-root trick.

\section{Background}

Our starting point is the staggered fermion formulation in four Euclidean dimensions.  The lattice Dirac equation is diagonalised by rewriting the spinor field as $\tilde{\psi}_n$, where
\[
\tilde{\psi}_n = T(n)^{\dagger} \psi_n
\]
and
\[
T(n) = \gamma_1^{n_1}\gamma_2^{n_2}\gamma_3^{n_3}\gamma_4^{n_4} \ .
\]

We retain only the first component of $\tilde{\psi}_n$, defining the field $\chi_n=\tilde{\psi}_n^1$.  The resulting massless action is
\[
S=\frac{1}{2}\sum_{n,\mu}\eta_{\mu}(n)\bar{\chi}_n[\chi_{n+\hat{\mu}} - \chi_{n-\hat{\mu}}] \ ,
\]
where 
\[
\begin{array}{lll}
\eta_{1}(n) = 1 & &  \\
\eta_{\mu}(n) = (-1)^{n_1+..+n_{\mu-1}} & : & \mbox{ for $\mu \not= 1$} \ .
\end{array}
\]

The lattice sites are grouped into hypercubes of side twice the original lattice spacing, each containing sixteen sites.  We relabel the coordinate of the lattice site $n$ with the coordinate of the origin of the hypercube, $m$, and the location of the site within the hypercube, $s$.  The relationship between the four component vectors $n$, $m$ and $s$ is $n=m+s$.  Together the hypercubes yield a fermion field that is free from degeneracy in momentum space
\begin{equation}
\label{transform}
\hat{\psi}_m = \mathcal{N}_0 \sum_s T(s) \chi_m^s \ ,
\end{equation}
where $\mathcal{N}_0$ is a normalisation constant.  For $\mathcal{N}_0 = \frac{1}{\sqrt{2}}$, the action is 
\begin{equation}
\label{action}
S=\sum_{m, \mu} b^4 \bar{\hat{\psi}}_m\left[(\gamma_{\mu} \otimes I)\partial_{\mu} + \frac{b}{2}(\gamma_5 \otimes \gamma^*_{\mu}\gamma_5)\Box_{\mu}\right]\hat{\psi}_m \ ,
\end{equation}
where $A\otimes B$ is the product of the spin and flavour spaces and $b$ is twice the original lattice spacing.

$\hat{\psi}_m$ is now a four by four matrix containing four degenerate flavours.  The action mixes both the spin and flavour space and the only operator that anticommutes with both terms in the action is $\gamma_5 \otimes \gamma_5$, generating a limited form of chiral symmetry.

\section{Flavour projection}
In the continuum limit, the second term in equation (\ref{action}) goes to zero, we have that $\hat{\psi}_m \rightarrow \hat{\psi}(m)$ and the columns of $\hat{\psi}(m)$ come to represent separate flavours of the field.  Each column can be isolated with the operator $\mathbf{\hat{P}}^{(b)}$, where 
\[
\mathbf{\hat{P}}^{(b)} \hat{\psi}(m)  = \hat{\psi}(m) P^{(b)}
\]
and the projection matrix $P^{(b)}$ is of the form \
\[
\begin{array}{cclccl}
\mbox{diag}(1,0,0,0) &:& \mbox{for } b=1 \ , \qquad & \qquad \mbox{diag}(0,1,0,0) &:&  \mbox{for } b=2 \ ,\\
\mbox{diag}(0,0,1,0)  &:& \mbox{for } b=3 \ , \qquad & \qquad \mbox{diag}(0,0,0,1)  &:& \mbox{for } b=4  \ .
\end{array}
\]
$P^{(b)}$ is defined through 
\begin{equation}
\label{proj}
P^{(b)} = \frac{1}{4}\left(1+i\alpha_b\gamma_1\gamma_2\right) \left(1+\beta_b\gamma_1\gamma_2\gamma_3\gamma_4\right)
\end{equation}
with 
\[
\begin{array}{ccl} 
\alpha_b & = & (-1,+1,-1,+1)^T \\ 
\beta_b & = & (-1,-1,+1,+1)^T \ .
\end{array}
\]
We can expand the projection matrix to give
\begin{equation}
\label{gammas}
P^{(b)} = \frac{1}{4}\left(1+i\alpha_b\gamma_1\gamma_2+\beta_b\gamma_1\gamma_2\gamma_3\gamma_4-i\alpha_b\beta_b \gamma_3\gamma_4\right)
\end{equation}
and this is equal to 
\[
\begin{array}{ccl}
P^{(b)} & = & \frac{1}{4}\Big( \mbox{diag}(1,1,1,1) + \alpha_b \mbox{diag}(-1,1,-1,1) \\
& & + \beta_b\mbox{diag}(-1,-1,1,1) + \alpha_b\beta_b\mbox{diag}(1,-1,-1,1)\Big) \ .
\end{array}
\]
Writing
\[
\begin{array}{ll}
P_1 = \mbox{diag}(1,1,1,1) \ , \qquad & \qquad P_2  = \mbox{diag}(-1,1,-1,1) \ , \\
P_3 = \mbox{diag}(-1,-1,1,1) \ , \qquad & \ \qquad P_4 = \mbox{diag}(1,-1,-1,1) \ , \\
\end{array}
\]
we can see that $[D, \mathbf{\hat{P}^{(b)}}]=0$, where $D$ is the staggered Dirac operator, because
\[
\left[D\psi(m)\right]P_i = D\left[\psi(m)P_i\right] \ .
\] 

At finite lattice spacing, $b\not= 0$ and the second term in the staggered Dirac operator breaks this associativity.  Thus $[D, \mathbf{\hat{P}}] \not=0$.  However, if we could transform the $P_i$ in a manner consistent with the effect of $D$ on $\hat{\psi}_m$, we would be able to maintain $[D, \mathbf{\hat{P}^{(b)}}]=0$ at finite lattice spacing. 

To implement this strategy, we write $\hat{\psi}_m P^{(b)}$ as 
\[
\begin{array}{ccl}
\hat{\psi}_m P^{(b)} & = & \frac{1}{4}\hat{\psi}_m\left(P_1 +\alpha_b P_2 +\beta_b P_3 + \alpha_b\beta_b P_4\right) \\
& = & \frac{1}{4}\left(\hat{\Psi}^1_m +\alpha_b \hat{\Psi}^2_m +\beta_b \hat{\Psi}^3_m +\alpha_b\beta_b\hat{\Psi}^4_m\right) \ .
\end{array}
\]
We introduce three additional lattices, in parallel to the first, and on the $i$-th lattice, $L_i$, we place the field $\hat{\Psi}^i_m$.  Flavour projection now becomes a projection between the lattices and to describe it we introduce the following operator to map a field from lattice $L_i$ to lattice $L_j$
\[
\hat{T}_{ij} : L_i\rightarrow L_j \ .
\]
Writing the fermion field across all four lattices as $\hat{\Psi}_m=\sum_{i}\hat{\Psi}^i_m$ and using equation (\ref{proj}), the flavour projection operator can now be written as
\[
\begin{array}{ccl}
\mathbf{\hat{P}}^{(b)}\hat{\Psi}_m & = & \frac{1}{4}\Big(1+\beta_b\left(T_{31} - T_{13} +T_{42} - T_{24} \right)\Big)\\
 & & \times \Big(1+\alpha_b\left(T_{21} - T_{12} + T_{43} - T_{34}\right)\Big)\hat{\Psi}_m \ .
\end{array}
\] 
Application of $\mathbf{\hat{P}}^{(1)}\hat{\Psi}$ will leave column $i$ on $L_i$.  However, because $D$ mixes the spin and flavour spaces equally on all four lattice, the same degrees of freedom of $\hat{\psi}_m$ will be projected out, irrespective of the order in which we apply $D$ and $\mathbf{\hat{P}}^{(b)}$ to $\hat{\psi}_m$.  Thus we have $[\mathbf{\hat{P}}^{(i)}, D] = 0$. 

\section{Chiral symmetry}
With the components from only one flavour of field on each lattice, we can see that the residual $\gamma_5\otimes\gamma_5$ chiral generator is sufficient to isolate the chiral components.  For example, the application $P_R\mathbf{\hat{P}}^{(1)}\hat{\Psi}$ leaves the positive chirality components of the first column of $\hat{\psi}_m$ on $L_1$ and of the second column of $\hat{\psi}_m$ on $L_2$.  It also leaves the negative chirality components of the third column of $\hat{\psi}_m$ on $L_3$ and of the fourth column of $\hat{\psi}_m$ on $L_4$.  Because the chiral projection isolates the same components of $\hat{\Psi}^i_m$ on each lattice, it does not interfere with the flavour projection and so we have that $[P_{R/L}, \mathbf{\hat{P}}^{(b)}]=0$.  

The effect of the projection on the chiral anomaly can be understood by considering $\mathbf{\hat{P}}^{(1)}P_R\hat{\Psi}$.  The application of $P_R$ to $\hat{\Psi}_m$ projects out two positive and two negative chirality spinors on each lattice which ensure that the chiral current is conserved.  Writing $\mathbf{\hat{P}}^{(b)}$ as
\[
\begin{array}{ccl}
\mathbf{\hat{P}}^{(b)} & = & \mathbf{\hat{P}}^{(b)}_{\beta}\mathbf{\hat{P}}^{(b)}_{\alpha} \\
\mathbf{\hat{P}}^{(b)}_{\beta} & = & \frac{1}{2}\Big(1+\beta_b\left(T_{31} - T_{13} +T_{42} - T_{24} \right)\Big)\\
\mathbf{\hat{P}}^{(b)}_{\alpha} & = & \frac{1}{2}\Big(1+\alpha_b\left(T_{21} - T_{12} + T_{43} - T_{34}\right)\Big) \ ,
\end{array}
\] 
we can see that $\mathbf{\hat{P}}_{\alpha}^{(1)}$ removes one positive and one negative chirality spinor from each lattice, leaving one positive and one negative chirality spinor behind.  At this stage the chiral current is still conserved on each lattice.  The subsequent application of $\mathbf{\hat{P}}_{\beta}^{(1)}$, separates the remaining spinors, placing them on different lattices.  This results in a chiral current flowing between $L_1$ and $L_3$ and between $L_2$ and $L_4$ which manifests itself as an anomaly on each lattice.

\section{Gauging the links}
The gauging of staggered fermion fields takes place in the $\chi_n$ basis.  However, the modified form of $\hat{\psi}_m$ that we place on each lattice leads us to introduce modified forms of $\chi_m^s$.  

Using the identity $Tr(T(n)^{\dagger}T(m)) = 4\delta^{n,m}$, the inverse of equation (\ref{transform}) can be shown to be 
\[
\chi_m^s = \frac{1}{4\mathcal{N}_0} Tr\left(T(s)^{\dagger} \hat{\psi}_m\right) \ .
\]
We define the modified field for lattice $i$ as
\begin{equation}
\label{invtrans}
\omega_m^{is} = \frac{1}{4\mathcal{N}_0} Tr\left(T(s)^{\dagger} \hat{\Psi}_m^i \right) \ .
\end{equation}

By writing the diagonal matrices in terms of the defining $\gamma$-matrix combinations from equation (\ref{gammas}), we can relate $\omega_m^{is}$ to $\chi^s_m$ as follows.
\begin{equation}
\label{chibas}
\begin{array}{cllll}
\omega_m^{1s} & = & \frac{1}{4\mathcal{N}_0} Tr\left(T(s)^{\dagger} \hat{\psi}_m \right)& = & \chi_m^s \\
\omega_m^{2s} & = &  \frac{i}{4\mathcal{N}_0} Tr\left(T(s)^{\dagger} \hat{\psi}_m \gamma_1\gamma_2\right) & = & -i\rho_{(s,12)}\chi_m^{\mathcal{C}_{12}s}\\
\omega_m^{3s} & = &\frac{1}{4\mathcal{N}_0} Tr\left(T(s)^{\dagger} \hat{\psi}_m \gamma_1\gamma_2\gamma_3\gamma_4 \right) & = & \rho_{(s,1234)}\chi_m^{\mathcal{C}_{1234}s}\\
\omega_m^{4s} & = & - \frac{i}{4\mathcal{N}_0} Tr\left(T(s)^{\dagger} \hat{\psi}_m \gamma_3 \gamma_4 \right) & = & i \rho_{(s,34)}\chi_m^{\mathcal{C}_{34}s} \ .
\end{array}
\end{equation}
Here $\rho_{(M,N)}$ is defined to be $(-1)^\nu$, where $\nu$ is the number of pairs $(m,n)$ with $m\in M$, $n\in N$ and $m>n$ \cite{becher}. We have also introduced $\mathcal{C}$ as a complementarity operator.  $\mathcal{C}_{12}s$ gives a vector that complements $s$ in the $\{1,2\}$ subspace, but is the same as $s$ in the $\{3,4\}$ subspace.  Similarly, $\mathcal{C}_{34}s$ gives a vector that complements $s$ in the $\{3,4\}$ subspace, but is the same in the $\{1,2\}$ subspace.  $\mathcal{C}_{1234}s$ gives a vector that complements $s$ in all four dimensions.  

On each lattice $\omega_m^{is}$ is gauged in the same manner as $\chi_m^s$.  Because of the linearity of the transform in equation (\ref{invtrans}), we can apply the projection operator $\mathbf{\hat{P}}^{(b)}$ to $\omega=\sum_{ims}\omega_{m}^{is}$ to project a single gauged spinor onto each lattice.

\section{Path integral formulation}
The path integral contains contributions from all four lattices. On each lattice, the path integral takes the form
\begin{equation}
\label{pi}
\int [dA]\left(\int [d\bar{\hat{\Psi}}^i][d\hat{\Psi}^i]e^{-\sum_{xx\prime}Tr\left(\bar{\hat{\Psi}}^i_{x\prime}K_{x\prime x}(A)\hat{\Psi}^i_{x} \right)} \right)^{L_0} \ ,
\end{equation}
where $L_0$ is a constant yet to be determined.  The fields on each lattice share the same degrees of freedom, but, because each lattice lies in a separate space, we can separately evaluate the contribution from each.  

The number of elements of  $\hat{\Psi}^2$, $\hat{\Psi}^3$ and $\hat{\Psi}^4$ that are the negative of their counterparts in $\hat{\Psi}^1$ is even and this gives the following equivalence
\[
[d\bar{\hat{\Psi}}^i][d\hat{\Psi}^i] = [d\bar{\hat{\Psi}}^1][d\hat{\Psi}^1] = [d\bar{\hat{\psi}}][d\hat{\psi}] \ ,
\]
giving us the same fermionic measure of integration on each lattice.  

If we consider the action on each lattice written as a matrix equation, it takes the form $S^i = \bar{\hat{\Psi}}^i_{\alpha} K_{\alpha\beta}(A)\hat{\Psi}^i_{\beta}$, where $K_{\alpha\beta}$ is the gauged staggered Dirac operator.  By rewriting the action as
\[
\bar{\hat{\Psi}}^i_{\alpha} K_{\alpha\beta}(A)\hat{\Psi}^i_{\beta} = \bar{\hat{\Psi}}^1_{\alpha} K^i_{\alpha\beta}(A)\hat{\Psi}^1_{\beta} = \bar{\hat{\psi}}_{\alpha} K^i_{\alpha\beta}(A)\hat{\psi}_{\beta} \ ,
\]
we introduce $K^i_{\alpha\beta}(A)$ in which an even number of columns are equal to the negative of their counterpart in $K_{\alpha\beta}(A)$.  Evaluating
\[
\left(\int [d\bar{\hat{\psi}}][d\hat{\psi}]e^{-\sum_{\alpha\beta}\bar{\hat{\psi}}_{\alpha}K^1_{\alpha\beta}(A)\hat{\psi}_{\beta}} \right)^{L_0}
\]
gives 
\[
\left(det[K^1(A)]\right)^{L_0} 
\]
which is equal to $\left(det[K(A)]\right)^{L_0}$.  Because $K^2_{\alpha\beta}(A)$, $K^3_{\alpha\beta}(A)$ and $K^4_{\alpha\beta}(A)$ differ from $K^1_{\alpha\beta}(A)$ by an even number of columns multiplied by $-1$, we have that 
\begin{equation}
\label{det}
\begin{array}{ccl}
\left(\int [d\bar{\hat{\psi}}][d\hat{\psi}]e^{-\sum_{\alpha\beta}\bar{\hat{\psi}}_{\alpha}K^i_{\alpha\beta}(A)\hat{\psi}_{\beta}} \right)^{L_0} & = & \left( det[K^i(A)]\right)^{L_0} \\
& = & \left(det[K^1(A)]\right)^{L_0} \\
& = & \left(det[K(A)]\right)^{L_0} \ .
\end{array}
\end{equation}
Hence, the products of all four contributions to the path integral is 
\[
\left(det[K(A)]\right)^{4L_0} \ .
\]
For the path integral to maintain the correct continuum limit, we require that $L_0=\frac{1}{4}$.  This gives an expression on each lattice that is of the same form as that used in the quarter-root trick (equation (\ref{det})).

If we evaluate the path integral in the $\omega$ basis, we find from equation (\ref{chibas}) that, for each site $m$ on each lattice $i$, the sixteen components of $\omega_m^{is}$ contain an even number of terms that are the negative of their counterpart amongst $\chi_m^s$.  Hence, we have that
\[
[d\bar{\omega}^i][d\omega^i] = [d\bar{\omega}^1][d\omega^1] = [d\bar{\chi}][d\chi] \ .
\]
As in the $\hat{\psi}$ basis, the $-1$ factors within the action can be incorporated into the Dirac operator to show that the path integral evaluates to $(det[M(A)])^{L_0}$ on each lattice, where $M(A)$ is the Dirac operator in this basis. 

If, instead of immediately evaluating the path integral on each lattice, we first apply $\mathbf{\hat{P}}^{(b)}$ to equation (\ref{pi}), the flavour projection will affect both the action and fermionic measure of integration.  As discussed, on each lattice, it will leave the action acting on a single flavour of field.  It will also project between the measures of integration on each lattice, to leave the components belonging to the same single flavour as in the action.   The resulting path integral on each lattice, describes a form of single gauged fermion.
 
\section{Conclusion}
In this paper, we have shown that it is possible to describe flavour projection in staggered fermions as a projection between four lattices where modified forms of the fermion field occupy each lattice.  The staggered Dirac operator respects this flavour projection because it mixes the spin and flavour spaces of the fields on each lattice equally.  We saw that, after flavour projection, the $\gamma_5 \otimes \gamma_5$ chirality of the staggered fermion formulation is sufficient to isolate the chiral components of the remaining fields.  We also showed that this system could be gauged in the usual way and that it was possible to project out a single gauged field.  We showed that the projection could be applied to the full path integral and that, prior to flavour projection, evaluating the path integral lead to a form consistent with the quarter-root trick commonly seen in lattice QCD.  

It is our hope that this work will contribute to the discussion on fermion doubling and that, by finding a natural home for the quarter-root trick, this formulation can contribute to the current debate on its validity.

\end{document}